\documentclass[preprint2]{aastex}

\begin{document}

\title{TIME DILATION AND QUASAR VARIABILITY}
\author{M. R. S. Hawkins}
\affil{University of Edinburgh, Royal Observatory, Blackford Hill,
Edinburgh EH9 3HJ, Scotland, UK}
\email{mrsh@roe.ac.uk}

\begin{abstract}
The timescale of quasar variability is widely expected to show the
effects of time dilation.  In this paper we analyse the Fourier power
spectra of a large sample of quasar light curves to look for such an
effect.  We find that the timescale of quasar variation does not
increase with redshift as required by time dilation.  Possible
explanations of this result all conflict with widely held consensus
in the scientific community.
\end{abstract}


\section{INTRODUCTION}

Time dilation is a fundamental property of an expanding universe.
In fact the increase of timescale by a factor of $(1+z)$ represents
a basic link between redshift and time which is essentially
related to the definition of time and is independant of cosmological
model parameters.  As a consequence, time dilation has generally
been assumed to be a property of the Universe even though it has
proved hard to measure directly.  Recently there has been new interest
in time dilation as a result of experiments where its effect is
large and must be taken into account.

There have been a number of claims by groups working on gamma ray
bursters \citep{m98} that time dilation is seen in the stretching of
peak-to-peak timescales.  This has then been used to support the
argument that the bursts are at cosmological distances.  It is not
clear however that the argument can be inverted to provide convincing
evidence for the existence of time dilation.

A more direct observation of time dilation has come from the
measurement of the decay time of distant supernova light curves
and spectra \citep{l96,g97,r97}.  Here one can make a good case that
the rest frame timescale is known, and hence directly detect any
time dilation effect at high redshift.  The results so far published
are very convincing, and strongly imply that time dilation has been
observed.

Another situation where one would expect to see a time dilation effect
is in the light curves of quasars, which are at cosmological distances
and vary on a timescale of years.  A number of groups have looked for
time dilation in quasar light curves \citep{h94,c96,h96}, but
so far it seems fair to say that no convincing detection has been made,
although one might argue that the sample sizes and data analysis
procedures were not adequate to detect the effect of time dilation
if it were present.

\section{POWER SPECTRUM ANALYSIS}

In order to measure time dilation in quasar light curves it is
necessary to find a way of characterising the timescale of variation.
The most popular parameterisation to date
has been the structure function, and several groups have measured
it for samples of light curves \citep{h94,c96}, as well as predicting
its shape for various models of quasar variability \citep{k98}.  The
autocorrelation function, which is closely related to the structure
function, has also been used \citep{h96}.  The main drawback of these
functions is that the points are not independent of each other, which
causes difficulties with error analysis, and makes them very hard to
interpret.

Fourier power spectrum analysis has not been used much in the analysis
of quasar light curves, probably because it requires a long run of
evenly spaced data to be effective.  However, given such a dataset
it provides some significant advantages over other methods, perhaps
the most important of which is the relative ease with which it may
be interpreted.  In this paper we apply it to a large sample of
quasars which have been homogeneously monitored every year for 24
years \citep{h96}.

The survey is based on a long series of UK 1.2m Schmidt plates of the
ESO/SERC field 287 centred on 21h 38m, -45$^{\circ}$.  The plates were
taken over a variety of timescales from hours to years, and in various
passbands including $B$, $R$ and $U$.  Of particular relevance to this
paper is a regular yearly monitoring of the field in the $B_{J}$
passband (Kodak IIIa-J emulsion with a Schott GG395 filter) from 1977
till 2000.  For most years four plates were obtained, but in a few
cases it was only possible to obtain one. A similar series of 18
yearly measures was also obtained in $R$, from 1983 till 2000.  The
plates were measured by the COSMOS or SuperCOSMOS machines at the
University of Edinburgh to give a catalogue of photometric measures
for some 200,000 objects in the central 19 square degrees of the field.
The photometric error on an observation from a single plate is about
0.08 mag.  When 4 plates were available the mean magnitude was used
giving a photometric error of about 0.04 mag.  More details of the
reduction procedure and error analysis are given by \citet{h96} and
references therein.

The quasars in the field were found by a variety of techniques,
including ultra-violet excess, variability, blue drop-out and
objective prism.   Altogether some 600 quasars have now been
identified, with confirming redshifts in the range $0.1 < z < 3.5$.
There are sufficient numbers that the quasars can be binned in both
redshift and luminosity to avoid the well-known degeneracy between
these two parameters.  All the quasars used in this study fluctuated
significantly in brightness over the 24 year monitoring period, with
an amplitude of mode 0.6 mag and a tail extending to 2 mag.  In order
to compare the spectrum of variations of subsamples of quasars from
the survey, a Fourier power spectrum was calculated for each light
curve.  The quasars were then binned in redshift and luminosity, in
such a way that each bin contained approximately 100 objects, with a
total of 407 quasars used for the analysis.

For the study of time dilation we first make the assumption that the
light variations are intrinsic to the quasars, and so the light curves
are subject to the effects of time dilation.  Thus for each light
curve we rescale the time interval by a factor of $(1+z)^{-1}$ where
$z$ is the redshift of the quasar, and re-sample the power spectra on
a uniform scale.  This should remove the effects of time dilation,
with the result that there should be no trend of timescale with
redshift.  This has the effect of shifting the contributions of all
quasars to higher frequencies.  It is a big effect, especially for
high redshift quasars, in several cases resulting in low frequency
bins being completely emptied.
 
\begin{figure*}
\epsscale{2.0}
\plotone{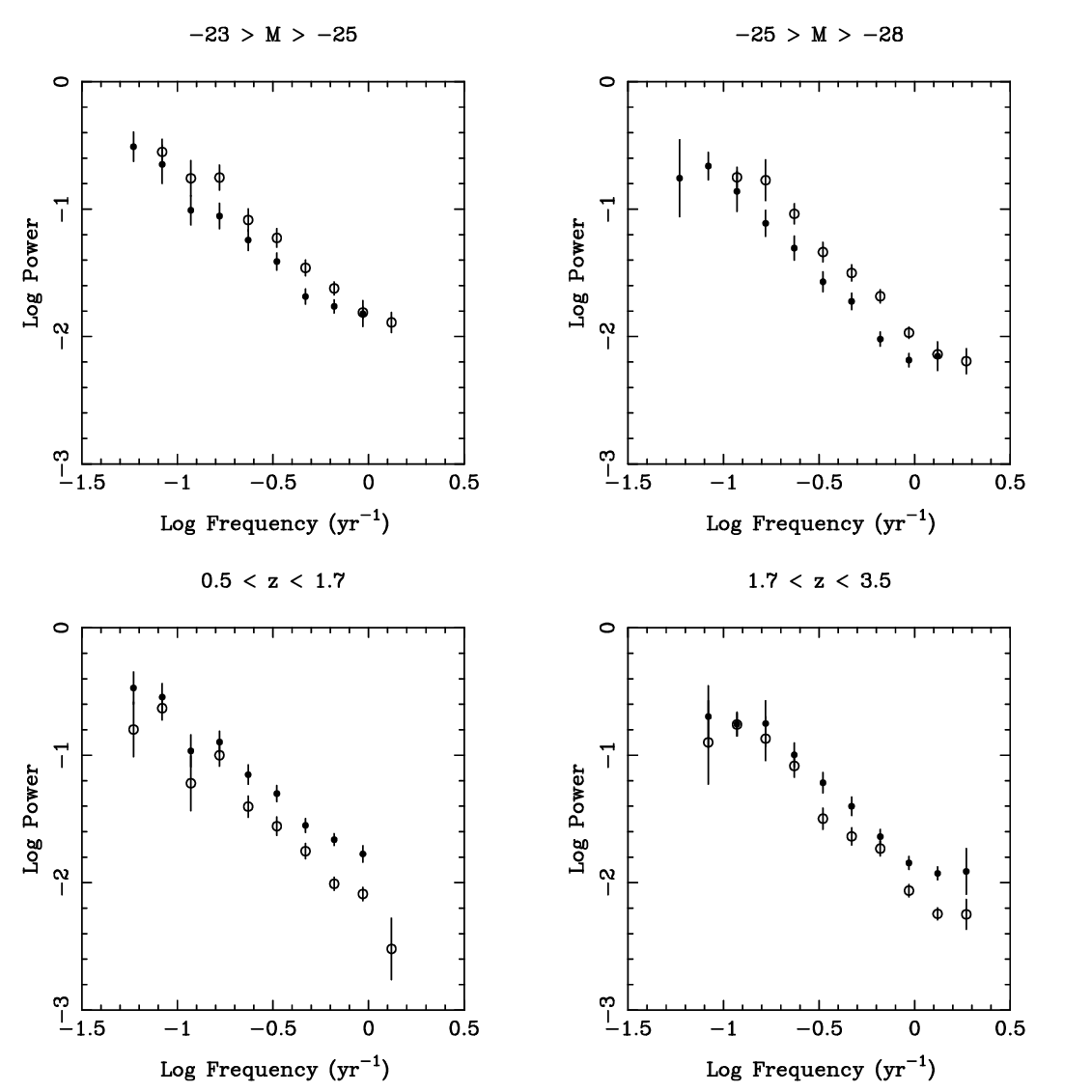}
\caption
{Fourier power spectra for sub-samples of quasar light curves
 in the quasar rest frame, with time dilation effects removed.
 The top two panels show data for low and high luminosity quasars.
 The filled and open circles are power spectra for low and high
 redshift objects respectively.  The bottom two panels show data for
 low and high redshift quasars.  The filled and open circles are power
 spectra for low and high luminosity objects respectively.
 \label{fig1}}
\end{figure*}

Fig.~\ref{fig1} shows the power spectra of samples of quasar light
curves, binned according to redshift and luminosity.  Each point
represents the average of value of all the contributions to that
frequency interval.  The top two panels show results for two
luminosity bins, with power spectra for low and high redshift quasars
being represented by filled and open circles respectively.  If the
timescales of the quasar light curves were subject to time dilation
one would expect no displacement between the two curves.  In  fact,
a $\chi^{2}$ test shows that for both luminosity bins the power
spectra are not coincident at the 99\% level.  The two power spectra
are separated by about 0.15 in the log in both luminosity bins.  This
is close to the offset produced by allowing for a $(1+z)$ scale change,
on the basis of the mean redshifts of the bins.  There is also some
indication that the power spectra have moved horizontally rather than
vertically from the morphology of the distributions.  The bottom two
panels show similar data for two redshift bins.  In this case low and
high luminosity quasars are represented by closed and open circles
respectively.  The mean redshift for the high and low luminosity
data in each plot only differs by about 0.03 in the log, and so
the removal of the $(1+z)$ factor should make little difference.
In fact the two luminosity bins are well separated, implying more
power or shorter timescales for low luminosity objects.
 
\begin{figure*}
\plotone{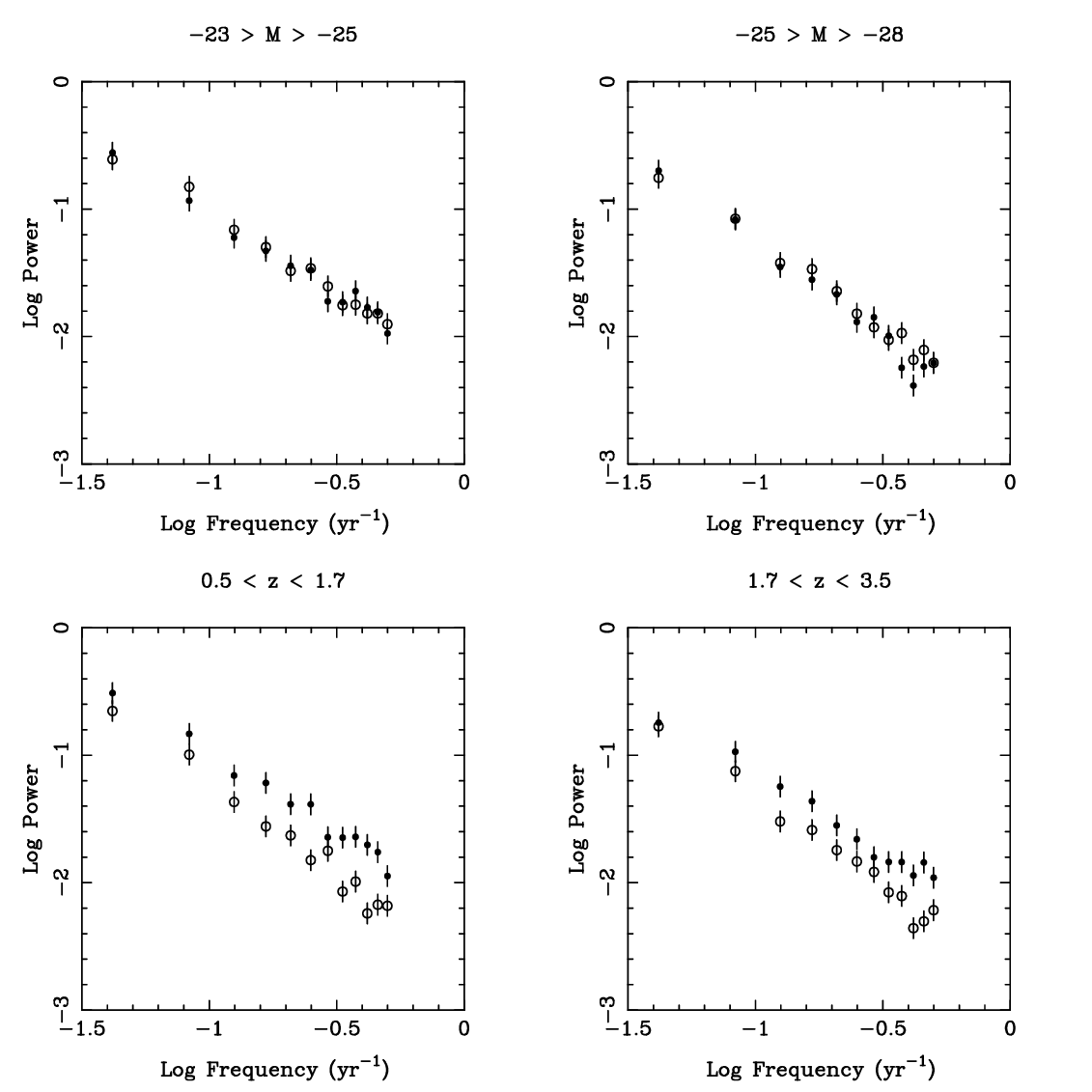}
\caption
{Fourier power spectra for sub-samples of quasar light curves
 in the observer's frame, with no allowance for time dilation.  The
 bins are as for Fig.~\ref{fig1}.  This figure shows that the
 observed timescale of quasar variation does not change with redshift
 for both luminosity bins.  It also shows that low luminosity quasars
 have more short timescale power than more luminous ones.
 \label{fig2}}
\end{figure*}
 
\begin{figure*}[t]
\begin{picture}(240,240)(-120,0)
\plotone{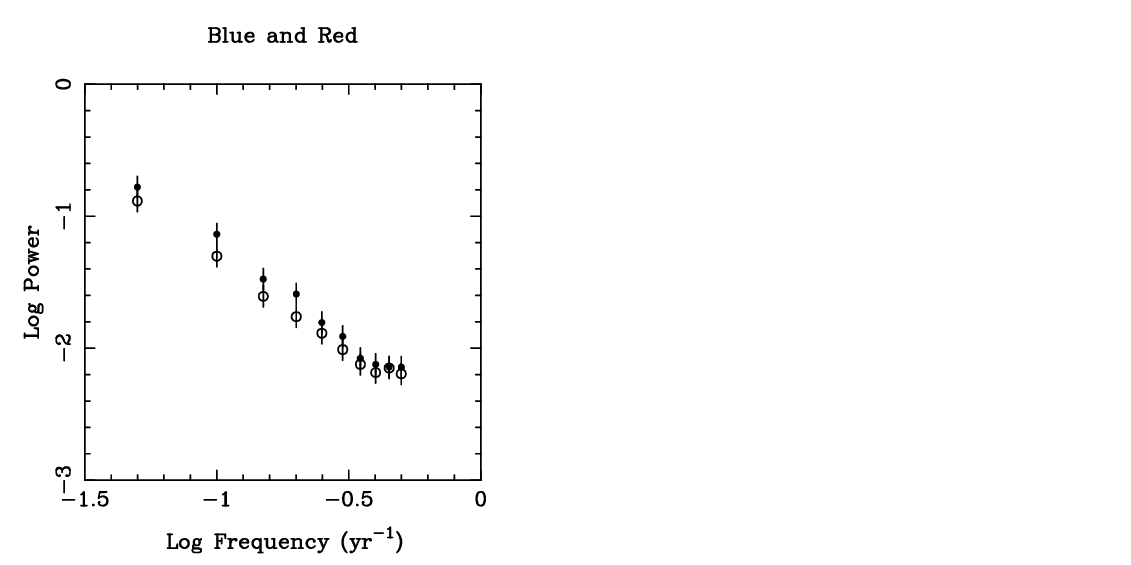}
\end{picture}
\caption
{Fourier power spectra for a sample of quasars in blue (closed
 circles) and red (open circles) passbands, in the observer's frame.
 \label{fig3}}
\end{figure*}

We now make no assumptions about the nature of the quasar variability,
but carry out the power spectrum analysis in the observer's
reference frame.  Fig.~\ref{fig2} shows power spectra of quasar
light curves as for Fig.~\ref{fig1} but this time with no
correction for time dilation.  For the two luminosity bins in the
top two panels it will be seen that all spectra show well-defined
linear (power law) relations.  On the basis of the mean redshift for
each of the bins, the effects of time dilation should result in a
horizontal offset of 0.15 between the power spectra in each of the
two top panels.  In fact, in each case the high and low redshift
data are superimposed, showing no change of timescale with redshift.
This is confirmed by a $\chi^{2}$ test which shows both pairs of
power spectra compatible at the 30\% level.
The two redshift bins in the bottom panels again show well defined
power laws for the power spectra, and in this case it is clear that
low luminosity quasars have more power on shorter timescales.  This
effect can also be seen by comparing the slopes of the power spectra
for low and high luminosity quasars in the top two panels, and
confirms that the power spectra are consistently measuring changes
in timescale.

\section{DISCUSSION}

The implication from Figs.~\ref{fig1} and~\ref{fig2} that
quasars do not suffer the effects of time dilation is hard to avoid.
There are however two possible ways out.  If the timescale of quasar
variation were a function of wavelength in the sense that timescales
were shorter in bluer passbands then this might possibly exactly
offset the effect of time dilation.  However, this can be directly
tested \citep{ht97}.  Fig.~\ref{fig3} shows power spectra for a sample
of about 200 quasar light curves in blue and red passbands.  If there
were a correlation of timescale with wavelength of the kind described
above, then the two power spectra in Fig.~\ref{fig3} should be
separated by 0.18 on the log scale, corresponding to the effective
wavelengths of 436 nm and 665 nm for the blue and red passbands.
In fact, the data for the blue light curves appear to be
systematically offset from the red by 0.06 in the log, but a $\chi^{2}$
test shows no significant difference at the 10\% level.  The only other
alternative is that the timescale of quasar variation decreases by a
factor $(1+z)$ towards high redshift by some as yet unspecified
physical process, to exactly cancel out the time dilation effect.  Such
a `cosmic conspiracy' has no independent motivation, and would require
a considerable degree of fine tuning.  In fact the shape of the plots
in Fig.~\ref{fig1} would appear to make such fine tuning an unrealistic
possibility.

There would appear to be three possible explanations for the lack of
a time dilation effect in quasar light curves, all of which conflict
with broad consensus in the astronomical community.  Firstly, time
dilation might not in fact be a property of the Universe, which would
effectively mean that the Universe was not expanding.  Apart from the
overwhelming support for the big bang theory, the direct measurements
of time dilation quoted above strongly argue against this.  The second
possibilty is that quasars are not at cosmological distances.  This is
an argument which was hotly disputed in the 1970s, with an emerging
consensus favouring cosmological distances.  This has subsequently been
strongly confirmed by studies of quasar host galaxies at high redshift.
The third possibility is that the observed variations are not intrinsic
to the quasars but caused by some intervening process at lower
redshift, such as gravitational microlensing.  Although this idea has
been strongly argued \citep{h96}, there is an opposing view that
variations in quasars are dominated by instabilities in the central
accretion disc.  The reality of this mode of variability in active
galactic nuclei is supported by detailed observations of Seyfert
galaxies \citep{p99} and gravitationally lensed quasars \citep{k97},
where the presence of intrinsic variations cannot be in doubt.  The
debate centres on whether this mechanism is responsible for the long
timescale large amplitude variations which dominate the power spectra
discussed in this paper.

\section{CONCLUSIONS}

Taking the various arguments outlined above at face value, and
accepting the case against microlensing, there does not appear to be a
satisfactory explanation for the absence of a time dilation effect in
quasar power spectra.  The arguments resting on an expanding Universe
and cosmological distances for quasars seem beyond challenge.  The
argument against microlensing is not so secure.  Apart from the
statistical evidence from quasar light curves \citep{h96},
microlensing has been unambiguously shown to take place in
gravitationally lensed quasar systems \citep{p98}, and dominates at
long timescales.  If this were a general phenomenon in quasars at
cosmological distances then the apparent absence of a time dilation
effect in quasar light curves would be explained.

\acknowledgements

I thank Gerson Goldhaber for a valuable suggestion for the presentation
of the data.

\end{document}